\documentclass[draft,12pt]{article}
\usepackage{amsmath,amsfonts,palatino,amsthm,epsf}
\newcommand{\beq}{\begin{equation}}
\newcommand{\eeq}{\end{equation}}

\begin{document}

\title{Modified Einstein-Cartan Gravity and its Implications for Cosmology}
\author{Wei Lu
\thanks{New York, USA, email address: weiluphys@yahoo.com}
\\
\\
}
\maketitle

\begin{abstract}
We propose a modification of Einstein-Cartan gravity equations. The modified cosmology departs from the standard model of cosmology for small Hubble parameter. A characteristic Hubble scale $h_0$, which is intrinsically related to cosmological constant, marks the boundary between the validity domains of the standard model of cosmology and modified cosmology. For large Hubble parameter, the standard model of cosmology is restored. In the opposite limit of small Hubble parameter, which is the case for present epoch, Lorentz-violating effects would manifest themselves. One of the implications is that there may be no need to invoke dark matter to account for cosmological mass discrepancies.
\end{abstract}


\newpage

\section{Introduction}

The dark matter hypothesis states that there is a vast amount of unseen mass in the universe. An alternative to dark matter is the modification of Newtonian dynamics\cite{A3, FM} (MOND). It is a classical dynamics theory, which explains the mass discrepancies in galactic systems without resorting to dark matter. 

We propose a relativistic theory with  modification of Einstein-Cartan equations. The modified torsion explicitly breaks local Lorentz gauge symmetry\cite{K}, while preserving diffeomorphism invariance. MOND is recovered in weak field limit\footnote{See \cite{BE, ZFS} for two examples of different relativistic MONDian theories. See \cite{FM} for a comprehensive list.}. We apply our modified gravity theory to cosmology. Friedmann equations are updated and their implications are discussed. The deviation from the standard model of cosmology is noticeable when Hubble parameter becomes comparable to or less than a characteristic Hubble scale. 

\section{Gauge Theory of Gravity}
In de Sitter gauge theory of gravity\cite{DESI, BH}, gravitational gauge field can be written as a Clifford-valued 1-form\cite{WL1}
\begin{align}
&A = \frac{1}{l}e + \omega, \\
&e = e^{a}\gamma_{a} = e^{a}_{\mu}dx^{\mu}\gamma_{a}, \\
&\omega = \frac{1}{4}\omega^{ab}\gamma_{ab}= \frac{1}{4}\omega^{ab}_{\mu}dx^{\mu}\gamma_{ab},
\end{align}
where $e$ is vierbein, $\omega$ is spin connection,  $\mu,a,b = 0, 1, 2, 3$, $\omega^{ab}_{\mu} = - \omega^{ba}_{\mu}$, and $\gamma_{ab} \equiv \gamma_{a}\gamma_{b}$. Here we adopt the summation convention for repeated indices. Clifford algebra vectors $\gamma_{a}$ observe anticommutation relations
\begin{align}
&\{\gamma_{a}, \gamma_{b}\} \equiv \frac{1}{2}(\gamma_{a}\gamma_{b}+\gamma_{b}\gamma_{a})=\eta_{ab},
\end{align}
where $\eta_{ab}$  is of signature $(+, -, - , -)$. 

The constant $l$ is related to Minkowskian vacuum expectation value (VEV) of gravity gauge field
\begin{align}
\bar A = \frac{1}{l} \bar{e} + \bar{\omega} = \frac{1}{l} \delta_{\mu}^a dx^{\mu} \gamma_a.
\end{align}

Gravity curvature $2$-form is given by
\begin{align}
F = dA + A^2 = R + \frac{1}{l}T + \frac{1}{l^2}e^2,
\end{align}
where spin connection curvature $2$-form $R$ and torsion $2$-form $T$ are defined by 
\begin{align}
R &= d\omega + \omega^2 = \frac{1}{4}R^{ab}\gamma_{ab} = \frac{1}{4}(d\omega^{ab} + \eta_{cd}\omega^{ac}\omega^{db})\gamma_{ab}, \\
T &= de + \omega e + e \omega = T^a \gamma_{a} = (de^{a} + \eta_{bc}\omega^{ab}e^{c}) \gamma_{a}. \\
\end{align}
Here exterior $\wedge$ products between forms are implicitly assumed.

One can write down the action for general relativity as\cite{WL1}
\begin{align}
S_G &= \frac{c^4}{8\pi G}\int{\left\langle -ie^{2}F \right\rangle} \\
		&= \frac{c^4}{8\pi G}\int{\left\langle -ie^{2}(R + \frac{1}{l^2}e^2) \right\rangle} \\
		&= \frac{c^4}{8\pi G}\int{\left\langle -ie^{2}(R + \frac{\Lambda}{24}e^2) \right\rangle} \\
		&= \frac{c^4}{32\pi G}\int{\epsilon_{abcd}e^a e^b (R^{cd} + \frac{\Lambda}{6}e^{c}e^{d})}, \\
\end{align}
where  $\Lambda$ is cosmological constant
\begin{align}
\Lambda = \frac{24}{l^2}, \label{CC}
\end{align}
$c$ is speed of light, $G$ is Newton constant\footnote{See \cite{WL1} for how Newton constant $G$ is related to $l$ and VEV of gravity Higgs field.}, $i$ is Clifford unit pseudoscalar
\begin{align}
i = \gamma_0\gamma_1\gamma_2\gamma_3, 
\end{align}
and $\left\langle \cdots\right\rangle$ means Clifford scalar part of enclosed expression. The action of gravity is invariant under local Lorentz gauge transformations. 

Field equations are derived by varying total action
\begin{align}
S = S_G + S_M 
\end{align}
with gauge fields $e$ and $\omega$ independently, where $S_M$ is matter part of the action. The resulted Einstein-Cartan equations read
\begin{align}
&\frac{c^4}{8\pi G}(Re + eR + \frac{\Lambda}{6}e^3) = \mathbb{T}i, \label{EC1}\\
&\frac{c^4}{8\pi G}(Te - eT) = \frac{1}{2}\mathbb{S}i, \label{EC2}
\end{align}
where $\mathbb{T}$ is energy-momentum current 3-form, and $\mathbb{S}$ is spin current 3-form.

\section{Lorentz Violation and Modified Gravity Equations}

Local Lorentz gauge transformation is characterized by
\begin{align}
&\mathbb{R}_{L}(x) = e^{\frac{1}{2}\epsilon^{ab}(x)\gamma_{ab}},
\end{align}
where $a,b = 0, 1, 2, 3$, $\epsilon^{ab}(x) = -\epsilon^{ba}(x)$, and $\gamma_{ab}$ are generators of Lorentz algebra. Gauge field $1$-form $e(x)$, spin connection curvature $2$-form $R(x)$, and torsion $2$-form $T(x)$ transform as\footnote{See e.g. chapter IX.7 of \cite{ZEE} for discussions about local Lorentz gauge transformation in the context of differential forms.}
\begin{align}
&V(x)\quad\rightarrow\quad  {\mathbb{R}_{L}(x)}V(x){\mathbb{R}_{L}(x)}^{-1}, \label{LOR1}
\end{align}
while spin connection $1$-form $\omega$ transforms differently as
\begin{align}
&\omega\quad\rightarrow\quad  {\mathbb{R}_{L}(x)}\omega{\mathbb{R}_{L}(x)}^{-1} - d{\mathbb{R}_{L}(x)}{\mathbb{R}_{L}(x)}^{-1}. 
\end{align}
The Einstein-Cartan equations \eqref{EC1} and \eqref{EC2} are covariant under local Lorentz gauge transformation, thanks to above transformation property for $e(x)$, $R(x)$, and $T(x)$. 

With the assumption of Lorentz symmetry violation, we study the remaining symmetry under local gauge transformation
\begin{align}
&\mathbb{R}_{S}(x) = e^{\frac{1}{2}\epsilon^{jk}(x)\gamma_{jk}}, \label{SP}
\end{align}
where $j,k=1,2,3$. Gravity gauge fields
\begin{align}
&e_{S} = e^{j}\gamma_{j} = e^{j}_{\mu}dx^{\mu}\gamma_{j}, \\
&e_{T} = e^{0}\gamma_{0} = e^{0}_{\mu}dx^{\mu}\gamma_{0}, \\
&\omega_{T} =\frac{1}{4}(\omega^{j0}\gamma_{j0} + \omega^{0j}\gamma_{0j})  = \frac{1}{2}\omega^{j0}_{\mu}dx^{\mu}\gamma_{j0},
\end{align} 
transform as
\begin{align}
&V(x)\quad\rightarrow\quad  {\mathbb{R}_{S}(x)}V(x){\mathbb{R}_{S}(x)}^{-1}, \label{SP1} 
\end{align}
while gauge field
\begin{align}
&\omega_{S} =\frac{1}{4}\omega^{jk}\gamma_{jk}
\end{align} 
transforms differently as
\begin{align}
&\omega_S(x)\quad\rightarrow\quad  {\mathbb{R}_{S}(x)}\omega_S(x){\mathbb{R}_{S}(x)}^{-1} - d{\mathbb{R}_{S}(x)}{\mathbb{R}_{S}(x)}^{-1}.  \label{SP2} 
\end{align}

With violation of Lorentz symmetry, we propose a change to Einstein-Cartan equation \eqref{EC2} in the form:
\begin{align}
&\frac{c^4}{8\pi G}(\tilde{T}e - e\tilde{T}) = \frac{1}{2}\mathbb{S}i. \label{EC3}
\end{align}
Here modified torsion $2$-form $\tilde{T}$ is defined by
\begin{align}
\tilde{T} &= T + \Delta T_{T} + \Delta T_{S},
\end{align}
where
\begin{align}
&\Delta T_{T} = (\alpha_T z)^{-\frac{1+\delta}{2}} (\omega_T e_S + e_S \omega_T), \\
&\Delta T_{S} = (\alpha_S z)^{-\frac{1+\delta}{2}} (\omega_T e_T + e_T \omega_T), \\
&z = \vert \frac{12(\frac{e}{l})^2(\omega_T \frac{e}{l} + \frac{e}{l} \omega_T)}{(\frac{e}{l})^4} \vert =  \vert \frac{12 l e^2(\omega_T e + e \omega_T)}{e^4} \vert.
\end{align}
The modified torsion $\tilde{T}$ breaks local Lorentz gauge symmetry, while preserving diffeomorphism invariance. Because of the transformation property \eqref{SP1} for $e_{S}$, $e_{T}$, and $\omega_{T}$, the modified Einstein-Cartan equations \eqref{EC1} and \eqref{EC3} are covariant under local gauge transformation \eqref{SP}\footnote{Since $\omega_{S}$ transforms differently as \eqref{SP2}, the modified torsion can not be dependent on $\omega_{S}$ individually.}. 

Three dimensionless parameters $\delta$, $\alpha_T$ and $\alpha_S$ are to be determined by comparing predictions of our proposal with astronomical observations. If $\alpha_T$ and $\alpha_S$ are equal, the modification to torsion can be written as
\begin{align}
\Delta T &= \Delta T_{T} + \Delta T_{S} = (\alpha z)^{-\frac{1+\delta}{2}} (\omega_T e + e \omega_T),
\end{align}
where $\alpha = \alpha_T = \alpha_S$.

\section{Weak Field Limit}
In static weak field  limit (gravity gauge field almost Minkowskian $A \approx \bar A  = \frac{1}{l} \delta_{\mu}^a \gamma_a dx^{\mu}$), the modified Einstein-Cartan field equations \eqref{EC1} and \eqref{EC3} are reduced\footnote{We are interested in galactic systems in this section. Spin current $\mathbb{S}$ and cosmological constant $\Lambda$ term are set to zero, since their effect is negligible.} to
\begin{align}
& \partial_i \omega_0^{i0} = \frac{4 \pi G}{c^2} \rho, \\
& \partial_i e_0^0 - \omega_0^{i0}(1 + (\alpha_T z)^{-\frac{1 + \delta}{2}}) = 0,
\end{align}
where 
\begin{align}
z = l (\omega_0^{i0}\omega_0^{i0})^{\frac{1}{2}}, 
\end{align}
and $\rho$ is mass density. 

The acceleration of a non-relativistic test body moving in the gravitational field is given by
\begin{align}
\vec{a} &= - c^2 \nabla e_0^0 = - \nabla V_N[1 +(\frac{|\nabla V_N|}{a_0})^{-\frac{1 + \delta}{2}}],
\end{align}
where 
\begin{align}
&\nabla^2 V_N =  c^2 \partial_i \omega_0^{i0} = 4 \pi G \rho,
\end{align}
and the characteristic acceleration scale $a_0$ is given by
\begin{align}
& a_0 = \frac{c^2}{\alpha_T l}. \label{A0}
\end{align}
It is intrinsically linked to cosmological constant \eqref{CC}  as
\begin{align}
& a_0 = \frac{c^2}{\alpha_T}\left(\frac{\Lambda}{24}\right)^{\frac{1}{2}}. \label{A0CC}
\end{align}
In the limit $|\nabla V_N| \gg a_0 $, Newtonian dynamics is restored, provided $1 + \delta > 0$. 
For $|\nabla V_N| \ll a_0 $, one can calculate circular orbit rotation velocity in potential 
\begin{align}
&V_N = -\frac{GM}{r} 
\end{align}
as
\begin{align}
v^4 &= a_0^{1+\delta}GM^{1-\delta} r^{2\delta}. \label{RV}
\end{align}
According to Tully-Fisher law\cite{A4} of galactic rotation curves, one has an estimation of parameter
\begin{align}
&\delta \approx 0. 
\end{align}
The characteristic acceleration is approximately
\begin{align}
&a_0 \approx 10^{-8}cm/s^2 \approx \frac{c^2}{6}(\frac{\Lambda}{3})^{\frac{1}{2}}.
\end{align}
Thus parameter $\alpha_T$ of our model is determined as
\begin{align}
\alpha_T = \frac{c^2}{la_0} = \frac{c^2}{a_0}(\frac{\Lambda}{24})^{\frac{1}{2}} \approx 2.
\end{align}

\section{Cosmology and Modified Friedmann Equations}
In this section, we apply our modification to cosmology\footnote{See \cite{DP, CL, NO, CFPS} for reviews of other modified gravity theories and their applications in cosmology. See \cite{KPM} for a review of challenges facing the standard model of cosmology.}. The spatially homogeneous and isotropic universe is described by Robertson-Walker (RW) metric
\begin{equation}
ds^2 = c^2dt^2 - a(t)^2\left(\frac{dr^2}{1-\kappa r^2/R_0^2} + r^2d\Omega^2\right),
\end{equation}
where $\Omega^2 = d\theta^2 + sin^2\theta d\phi^2$. With the above metric, \eqref{EC1} and \eqref{EC3} are reduced to modified Friedmann equations as 
\begin{align}
&\tilde{H}^2 = \frac{8 \pi G}{3} \rho + \frac{c^2}{3}\Lambda - \frac{\kappa c^2}{R_0^2 a^2}, \label{MFR} \\ 
&\frac{d(a\tilde{H})/dt}{a} = -\frac{4 \pi G}{3} (\rho + \frac{3}{c^2}p) + \frac{c^2}{3}\Lambda, \label{MAC}
\end{align}
where 
\begin{align}
&\tilde{H} \left( 1 + \left(\tilde{H}/h_0\right)^{-\frac{1+\delta}{2}} \right)= H, \label{MH} \\ 
&H = \frac{\dot{a}}{a} = \frac{da/dt}{a},\\
&h_0 = \frac{c}{3\alpha_Sl} = \frac{c}{3\alpha_S}\left(\frac{\Lambda}{24}\right)^{\frac{1}{2}}. \label{H0}
\end{align}
Here $H$ is Hubble parameter, $\tilde{H}$ is modified Hubble parameter, $h_0$ is a characteristic Hubble scale, $\rho$ is mass density, and $p$ is pressure. Spin current $\mathbb{S}$ is assumed to be zero. It is noted that torsion modification $\Delta T_{T}$ is relevant for Schwarzschild metric, while $\Delta T_{S}$ is relevant for RW metric. Hence, $a_0$ and $h_0$ are dependent on $\alpha_T$ and $\alpha_S$, respectively. The relation between the characteristic Hubble scale \eqref{H0} and MOND acceleration scale \eqref{A0} is
\begin{align}
&h_0 = \frac{1}{3c}\frac{\alpha_T}{\alpha_S}a_0. \label{H0A0}
\end{align}

Since the free parameter $\delta$ is estimated to be very close to zero, we will assume that $\delta = 0$ in the following analysis. The modified Hubble parameter $\tilde{H}$ is determined via equation \eqref{MH} as
\begin{align}
&\tilde{H} = \mu(H/h_0)H,
\end{align}
with interpolation function
\begin{align}
&\mu(x) \quad\rightarrow\quad 1 \quad \quad \quad \quad \quad  for \quad  x \gg 1, \label{IN1}\\
&\mu(x) \quad\rightarrow\quad x \quad \quad \quad \quad \quad for \quad  x \ll 1. \label{IN2}
\end{align}
One can potentially regard \eqref{MFR} as a phenomenological model
\begin{align}
&\left(\mu(H/h_0)H\right)^2 = \frac{8 \pi G}{3} \rho + \frac{c^2}{3}\Lambda - \frac{\kappa c^2}{R_0^2 a^2}, \label{MFR2}
\end{align}
with the interpolation function specified by \eqref{IN1} and \eqref{IN2}.

In the limit of $H \gg h_0 $, one has $\tilde{H} \simeq H $. Therefore, \eqref{MFR} and \eqref{MAC} are reduced to the usual Friedmann and acceleration equations\cite{COS1, COS2} as,
\begin{align}
&\left(\frac{\dot{a}}{a}\right)^2 = \frac{8 \pi G}{3} \rho + \frac{c^2}{3}\Lambda - \frac{\kappa c^2}{R_0^2 a^2}, \label{FR} \\ 
&\frac{\ddot{a}}{a} = -\frac{4 \pi G}{3} (\rho + \frac{3}{c^2}p) + \frac{c^2}{3}\Lambda, \label{AC}
\end{align}
where $\ddot{a} = d^2a/dt^2$. 

In the opposite limit of $H \ll h_0 $, $\tilde{H}$ is given by
\begin{align}
&\tilde{H} \simeq \frac{H}{h_0}H.
\end{align}
The modified Friedmann equations then read
\begin{align}
&\frac{1}{h_0^2}\left(\frac{\dot{a}}{a}\right)^4 = \frac{8 \pi G}{3} \rho + \frac{c^2}{3}\Lambda - \frac{\kappa c^2}{R_0^2 a^2}, \label{MFR1} \\
&\frac{1}{h_0}\left(\frac{2\dot{a}\ddot{a}}{a^2} - \frac{\dot{a}^3}{a^3}\right) = -\frac{4 \pi G}{3} (\rho + \frac{3}{c^2}p) + \frac{c^2}{3}\Lambda.
\end{align}

Let's study a simple case of one-component universe with $\kappa = 0$, $\Lambda = 0$, and matter density $\rho = \rho_0 a^{-3}$. We assume that it starts with $H \gg h_0 $. From equation \eqref{FR}, $\dot{a}$ follows 
\begin{align}
&\dot{a} \sim t^{-\frac{1}{3}},
\end{align}
which is decelerating. Eventually, the decreasing Hubble parameter will enter the regime $H \ll h_0$. Therefore, according to \eqref{MFR1}, $\dot{a}$ should follow
\begin{align}
&\dot{a} \sim t^{\frac{1}{3}},
\end{align}
which is accelerating. This scenario of late-time cosmic speed-up without cosmological constant is otherwise not possible in the standard model of cosmology\footnote{See \cite{PE, LLWW} for reviews of cosmological constant and its implications for cosmology. See e.g. \cite{DDG, CDTT} for earlier theories of cosmic acceleration without cosmological constant.}.

For certain values of $\kappa > 0$ and $\Lambda > 0$, numerical simulations show that the universe can even experience two periods of decelerating and accelerating phases. The first cycle of deceleration and acceleration is dominated by matter and characterized by $H \gg h_0$ and $H \ll h_0$, respectively. The second cycle is driven by positive curvature and cosmological constant, respectively. 

\section{Modified Density Parameter}
Dividing the new Friedmann equation \eqref{MFR} by $\tilde{H}^2$, one can get the density contributions for different components of the universe. The modified density parameter for baryonic matter is given by 
\begin{align}
&\tilde{\Omega}_b = \frac{8 \pi G}{3\tilde{H}^2} \rho_b = \frac{H^2}{\tilde{H}^2} \frac{8 \pi G}{3H^2} \rho_b  = \frac{H^2}{\tilde{H}^2}\Omega_b,
\end{align}
where $\Omega_b = \frac{8 \pi G}{3H^2} \rho_b$ is the usual density parameter for baryonic matter.

In the standard model of cosmology, cold dark matter(CDM) is invoked as an additional source of matter, since $\Omega_b$ is lower than what is observed. Here we propose that there is neither galactic CDM nor cosmological CDM. The modified density parameter $\tilde{\Omega}_b$ can be higher than the observed value $\Omega_b$, thanks to the factor $H^2/\tilde{H}^2$. This may eliminate the need for CDM.

Now we try to determine the magnitude of $H^2/\tilde{H}^2$. With the estimated value of characteristic acceleration \cite{FM} in terms of Hubble constant $H_0$ (which is the present value of Hubble parameter $H_0 = H|_{t=t_0}$),
\begin{align}
&a_0 \simeq \frac{c}{6} H_0,
\end{align}
equation \eqref{H0A0} gives
\begin{align}
&h_0 \simeq \frac{1}{18} \frac{\alpha_T}{\alpha_S} H_0. \label{H0H0}
\end{align}

With \eqref{H0H0} and \eqref{MH} (and $\delta = 0$), the present value of factor $(H^2/\tilde{H}^2)|_{t=t_0}$ can be calculated as $1.6$ or $6.9$ for $\alpha_T/\alpha_S = 1$ or $18$ (i.e. $H_0/h_0 = 18$ or $1$), respectively.

\section{Conclusion}
We propose a modification of Einstein-Cartan equations. Spin current is coupled to modified torsion, which breaks local Lorentz gauge symmetry and leaves diffeomorphism invariance intact. 

By setting the free dimensionless parameter $\delta$ to zero, one recovers MOND in weak field limit. Galactic rotation curves are explained without invoking dark matter. The characteristic acceleration scale $a_0$ is intrinsically linked to cosmological constant via VEV of gravity gauge field.

We then apply the new gravity theory to cosmology. The updated Friedmann equations are dependent on a modified Hubble parameter. The modified cosmology is in a sense similar to MOND: one replaces Hubble parameter $H$ with $\mu(H/h_0)H$ in Friedmann equation, whereas for MOND one replaces acceleration $a$ with $\mu(a/a_0)a$ in Newton equation. The deviation from the standard model of cosmology is noticeable when Hubble parameter becomes comparable to or less than $h_0$. The characteristic Hubble scale $h_0$ is proportional to MOND acceleration scale $a_0$.

One of the implications is that there may be no need to invoke dark matter to account for cosmological mass discrepancies. Another interesting observation is that our model can accommodate late-time cosmic acceleration without cosmological constant.

\section*{Acknowledgments}

I am grateful to Salvatore Capozziello, Friedrich Hehl, Arthur Kosowsky, Pavel Kroupa, Kenneth Macleod, Sergei Odintsov, Dirk Puetzfeld, and Tom Zlosnik for helpful correspondences.

\end{document}